\documentclass{Interspeech2024}

\interspeechcameraready

\usepackage{amsmath,graphicx}
\usepackage{siunitx}
\usepackage{amsmath}
\usepackage{amssymb}
\usepackage{multirow}
\usepackage[table,xcdraw]{xcolor}
\usepackage{multirow}
\usepackage{tikz}
\usepackage{tkz-graph}
\usetikzlibrary{positioning}
\usetikzlibrary{bayesnet}
\usetikzlibrary{matrix}
\usepackage{makecell}
\usepackage{booktabs}

\usepackage{subcaption}

\title{Audio Editing with Non-Rigid Text Prompts}

\name[affiliation={1,2}]{Francesco}{Paissan}
\name[affiliation={3,2}]{Luca}{Della Libera}
\name[affiliation={4}]{Zhepei}{Wang}
\name[affiliation={4}]{Paris}{Smaragdis}
\name[affiliation={3,2}]{Mirco}{Ravanelli}
\name[affiliation={5,3,2}]{Cem}{Subakan}

\address{
  $^1$Fondazione Bruno Kessler, Italy, $^2$Mila-Québec AI Institute, Canada $^3$Concordia University, Canada \\
  $^4$University of Illinois at Urbana-Champaign, USA $^5$Laval University, Canada
}
\email{fpaissan@fbk.eu, luca.dellalibera@mail.concordia.ca, zhepeiw2@illinois.edu, paris@illinois.edu, 	mirco.ravanelli@concordia.ca, cem.subakan@ift.ulaval.ca}


\begin{document}

\maketitle

\begin{abstract}
In this paper, we explore audio editing with non-rigid text prompts via Latent Diffusion Models. Our methodology is based on carrying out a fine-tuning step on the latent diffusion model, which increases the overall faithfulness of the generated edits to the input audio. We quantitatively and qualitatively show that our pipeline obtains results which outperform current state-of-the-art neural audio editing pipelines for addition, style transfer, and inpainting. Through a user study, we show that our method results in higher user preference compared to several baselines. We also show that the produced edits obtain better trade-offs in terms of fidelity to the text prompt and to the input audio compared to the baselines. Finally, we benchmark the impact of LoRA to improve editing speed while maintaining edits quality.
\end{abstract}
%

%
%
%

\section{Introduction}
Generative models have recently made incredible strides~\cite{yang2023diffusion, liu2024visual, li2024snapfusion, Karras2017ProgressiveGO}. In particular, text-prompted models, such as DALL-E \cite{Ramesh2021ZeroShotTG} and Stable Diffusion \cite{Rombach2021HighResolutionIS}, have shown impressive performance in text-prompted image generation. Parallel to generative modelling advancements, neural editing has also attracted attention, particularly in the computer vision literature \cite{Brooks2022InstructPix2PixLT, Couairon2022DiffEditDS}. Nonetheless, there has been a considerable amount of progress in the audio synthesis literature as well~\cite{huang2023make, Liu2023AudioLDM2L}, with the current state-of-the-art capable of generating convincing audio samples using text prompts. Examples include models such as AudioLDM \cite{liu2023audioldm}, which takes advantage of a joint text-audio latent space given by a CLAP (Contrastive Language-Audio Pretraining) \cite{clap} encoder. Other prominent examples include AudioGen \cite{kreuk2022audiogen} and MusicGen~\cite{copet2024simple}, which use a discretized latent space and inject text embeddings into an autoregressive prior in this space, and TANGO \cite{ghosal2023tango}, which directly injects pretrained large language model embeddings as conditioning for a latent diffusion process. Additionally, there exist {a few} models to create audio editing pipelines out of latent diffusion models. For example, AUDIT \cite{wang2023audit} enables editing audio using fixed commands such as ``Add'' a certain sound, ``Drop'' a sound excerpt, or ``Replace'' two sounds. Nonetheless, this approach limits the scope of the audio edits to a fixed set of commands. SDEdit~\cite{Meng2021SDEditGI}, instead, computes the audio using a two-step approach. First, it adds noise to the input, then denoises the resulting noisy audio through the SDE prior based on the text conditioning, thus breaking free of the limitations of AUDIT. Depending on the amount of noise injected, SDEdit can generate audio samples that are not faithful to the input audio, thus limiting its application scenarios.

Imagic \cite{kawar2023imagic} has recently been proposed to obtain free-form text-prompted edits from diffusion models. The intuition behind Imagic lies in interpolating the conditioning vectors, i.e. the text prompts. Similarly to SDEdit, Imagic allows for non-rigid image edits, as the approach does not require a closed set of commands and can be adapted to any new text prompt.

This paper explores the possibility of obtaining high-fidelity, non-rigid text-prompted edits for audio recordings. We emphasize that differently from existing audio generative models such as AudioLDM \cite{liu2023audioldm}, which is capable of style transfer, we aim to remain more faithful to the input audio, that is, to better preserve the onsets and offsets of audio events. Moreover, we work with 10-second audio signals, which leaves a large degree of freedom in the generated audio, possibly more than the images explored in the original Imagic paper. 

\noindent Our contributions in this paper are as follows:
\begin{itemize}
    \setlength\itemsep{.003cm}
    \item We propose to use a latent-space fine tuning and interpolation method to create non-rigid text-prompted audio edits that remain faithful to the input audio;
    \item We propose a quantitative evaluation method (sum of audio and text CLAP scores) for measuring perceptual semantic similarity to the text prompt and to the input audio, {and we show that our approach obtains a better overall tradeoff in terms of the proposed metric}; 
    \item We propose to use LoRA~\cite{Hu2021LoRALA} in the editing pipeline to significantly speed up the editing speed, while not sacrificing editing performance.
\end{itemize}

\tikzstyle{block} = [draw, fill=lightgray, rectangle, 
    minimum height=3em, minimum width=4em]
\tikzstyle{sumt}   = [circle, minimum width=8pt, draw, inner sep=0pt, path picture={\draw (path picture bounding box.east) -- (path picture bounding box.west) (path picture bounding box.south) -- (path picture bounding box.north);}]
\begin{figure*}
    \centering
    \resizebox{0.9\textwidth}{!}{
    \begin{tikzpicture}
            
            \node [draw, rectangle, minimum width=19.4cm, minimum height=3.9cm, very thick, fill=yellow!10] at (7.3, 0) (plateall) {};
            \node [draw, rectangle, minimum width=12.1cm, minimum height=3.5cm, very thick, fill=green!5] at (3.85, 0) (plate2) {};
            \node [] (inp) {\includegraphics[scale=0.25, trim=1cm 3cm 5cm 4cm, clip]{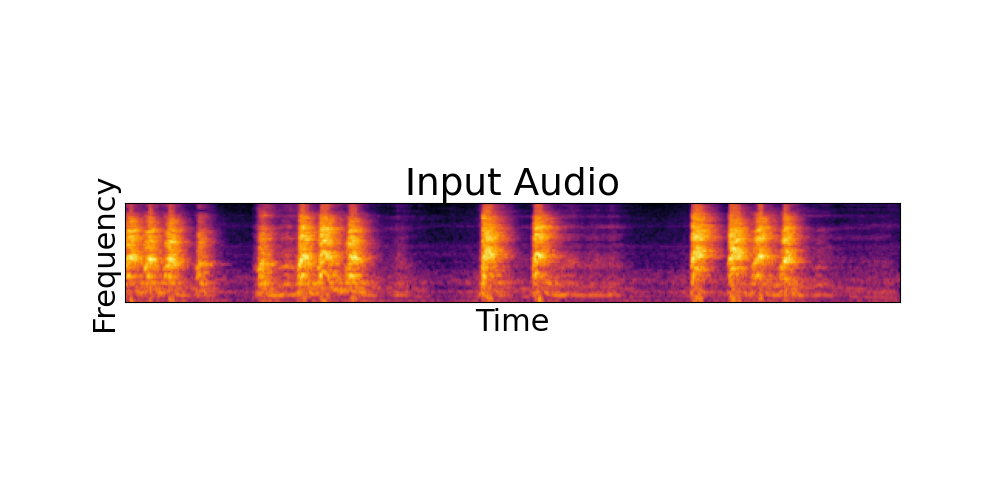}};
            \node [right of=inp, block, xshift=2.6cm, text width=1cm, align=center] (vaeencoder) {VAE\\Encoder};
            \node [right of=vaeencoder, draw, rectangle, minimum width=0.5cm, minimum height=2cm, xshift=0.4cm] (z) {$\mathbf z$};  
            \node [right of=z, sumt, xshift=0.1cm, scale=2] (sum) {};
            \node [above of=sum, yshift=.3cm] (noise) {\text{Noise}};
            \node [right of=sum, block, xshift=0.7cm, text width=1.5cm, align=center] (difmod) {Pretrained Diffusion Model};
            \node [above of=difmod, yshift=.2cm] (eopt) {$\mathbf e_\text{text}$};
            \node [right of=difmod, draw, rectangle, minimum width=0.5cm, minimum height=2cm, xshift=0.7cm] (out) {$\mathbf z'$};  
            \node [below of=difmod, yshift=-.5cm, label={[shift={(-.5,-.1)}]Reconstruction Loss (step 1+2)}] (nbelow) {};
            \node [right of=out, block, xshift=0.4cm, text width=1cm, align=center] (vaedec) {VAE\\Decoder};
            \node [right of=vaedec, xshift=2.5cm] (outaudio) {\includegraphics[scale=0.25, trim=2.1cm 3cm 5cm 4cm, clip]{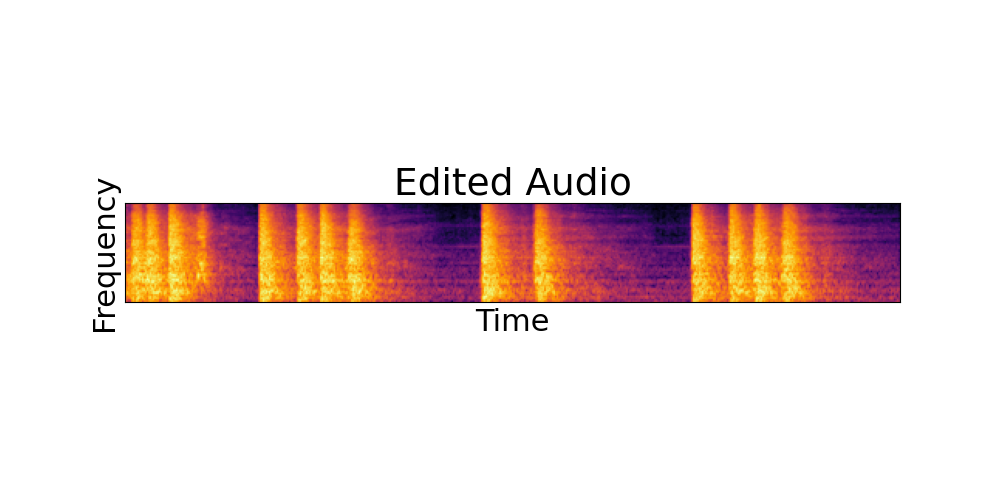}};
            \node [] at (3, 1.5) () {\textbf{\emph{Step 1+2}}};
            \node [] at (12.5, 1.7) () {\textbf{\emph{Step 3}}};

            \draw [->] (noise) -- (sum.north);
            \draw [->] (inp) -- (vaeencoder);
            \draw [->] (vaeencoder) -- (z);
            \draw [->] (sum) -- (difmod);
            \draw [->] (z) -- (sum);
            \draw [->] (difmod) -- (out);
            \draw [<->] (z.south) |- (nbelow.center) -| (out.south); 
            \draw [->] (eopt) -- (difmod);
            \draw [->] (out) -- (vaedec);
            \draw [->] (vaedec) -- (outaudio);
    \end{tikzpicture}
    }
    \caption{In step 1, we optimize with respect to $\mathbf e_\text{text}$ to minimize the reconstruction error in Eq.~\eqref{eq:difeq}, where $\mathbf z=\text{VAEEncoder}(X_\text{inp})$. The resulting optimized text embedding is denoted with $\mathbf e_\text{opt}$. In step 2, we optimize with respect to the Diffusion Model parameters to minimize the same reconstruction loss as in step 1. Note that in steps 1 and 2 only the part shown with the green box is used. In step 3, the text embedding is set as the linear combination of target embedding and the optimized embedding such that $\mathbf e_\text{text} = \eta \mathbf e_\text{target} + (1-\eta) \mathbf e_\text{opt}$. In step 3, the whole pipeline denoted by the yellow box is used.}
    \label{fig:model-pipeline}
\end{figure*}
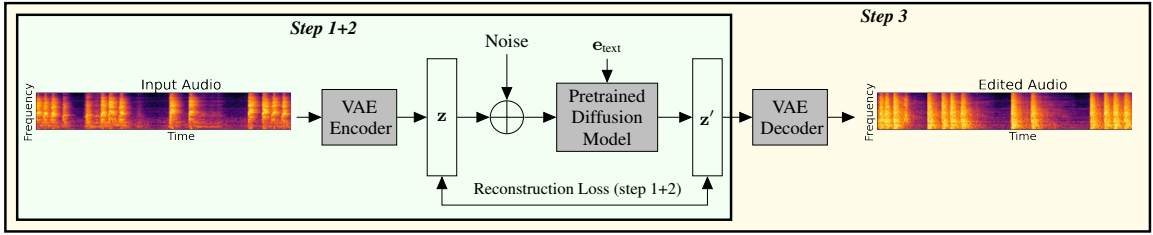

\section{Methodology}
\label{sec:methods}

The text-prompted audio generation pipeline we use in this paper is a latent diffusion model (LDM) \cite{Rombach2021HighResolutionIS}. The latent diffusion is applied to the latent space given by the Variational Autoencoder (VAE) \cite{Kingma2014} applied to mel-spectra. Starting from white noise and a text conditioning vector, the reverse latent diffusion process approximates the latent representation of the original noise into a latent representation that corresponds to the target audio. Afterwards, this representation is fed to the VAE decoder to create a mel-spectrogram and passed through a pretrained HiFi-GAN \cite{Kong2020HiFiGANGA} that works as a vocoder.

The editing method we propose in this paper is based on learning to reconstruct the latent representation of the original audio by optimizing the text conditioning and the diffusion model. This also translates into modifying the latent space of the diffusion model conditioning so that it is conceivable to interpolate between the original and the target audio just by interpolating the relative text embeddings on the learned manifold. For this purpose, we adapt the methodology introduced in \cite{kawar2023imagic} for an image generation model. The editing pipeline, depicted in Fig. \ref{fig:model-pipeline}, is built on three main stages: \emph{i) Embedding optimization ii) Fine-tuning of the LDM iii) Interpolation.}


\subsection{Step 1: Embedding optimization}
In the first stage of our method, similarly to Imagic \cite{kawar2023imagic}, the text embedding input $\mathbf e_\text{text}$ is optimized such that the following diffusion process error is minimized: 
\begin{align}
    \mathcal L(\mathbf z, \mathbf e_\text{text}, t) = \mathbb E_{t, \epsilon} [\epsilon - f_\theta (\mathbf z_t, t, \mathbf e_\text{text})], \label{eq:difeq}
\end{align}
Where $t$ denotes the number of timesteps, which is randomly selected from a uniform distribution between 1 and 1000, $f_\theta$ is the UNet \cite{ronneberger2015unet} employed to predict the noise, and $\mathbf z_t$ refers to a noisy version of the input latent representation $\mathbf z$.
Namely, the conditioning vector $\mathbf e_\text{text}$ is optimized such that the UNet $f_\theta$ approximates the total amount of noise $\epsilon$ that would have been added by the forward diffusion process asymptotically. In practice, we initialize the embedding $\mathbf e_\text{text}$ from the embedding of the target text prompt $\mathbf e_\text{target}$. We denote the result of this embedding optimization process with $\mathbf e_\text{opt}$. Note that effectively, as shown in Fig.~\ref{fig:model-pipeline}, estimation of this noise minimizes the error between the VAE encoder $\mathbf z$ output and the reconstruction $\mathbf z'$ given by the reverse diffusion process.

\subsection{Step 2: Fine-tuning}
The output of Step 1 might not perfectly reconstruct the original audio. Thus, this step further improves the ability of the model to reconstruct the input signal by optimizing the parameters of the UNet, $f_\theta$, used in the diffusion process. For this, the same loss function is used as in Eq.~\eqref{eq:difeq}. This means that at the end of the first two steps, the sample generated from the diffusion model, when conditioned with $\mathbf e_\text{opt}$ is the original audio. This is of crucial importance for learning to create edits that are faithful to the input by either containing the same acoustic events or the same identities.

\subsection{Step 3: Interpolation}
In this step, we interpolate between the target text embedding $\mathbf e_\text{target}$ and $\mathbf e_\text{opt}$, such that
\begin{equation}
    \label{eq:interpolate}
    \mathbf e_\text{int} : = \eta \mathbf e_\text{target} + (1-\eta) \mathbf e_\text{opt},
\end{equation}
where $\eta$ is a scalar between 0 and 1. That is, the text conditioning embedding $\mathbf e_\text{text}$ is set to $\mathbf e_\text{int}$ when generating the edit. After the diffusion fine-tuning and embedding optimization, for $\eta=0$ the diffusion process reconstructs the original audio samples. Instead, for $\eta$ values closer to 1, the edit strength will increase as the impact of the target embedding generated audio is higher. It is important to find a good value for $\eta$ such that the output of the editing model remains faithful to the input audio.

\tikzstyle{dictsmall} = [draw, thick, fill=white!10, rectangle, 
    minimum height=1.0cm, minimum width=5cm] 
\begin{figure*}[ht]
    \centering
    \resizebox{0.99\textwidth}{!}{
    \centering
    \begin{tikzpicture}[auto] 
        \node [draw=none, fill=none] (ex1)  { \includegraphics[width=0.25\textwidth]{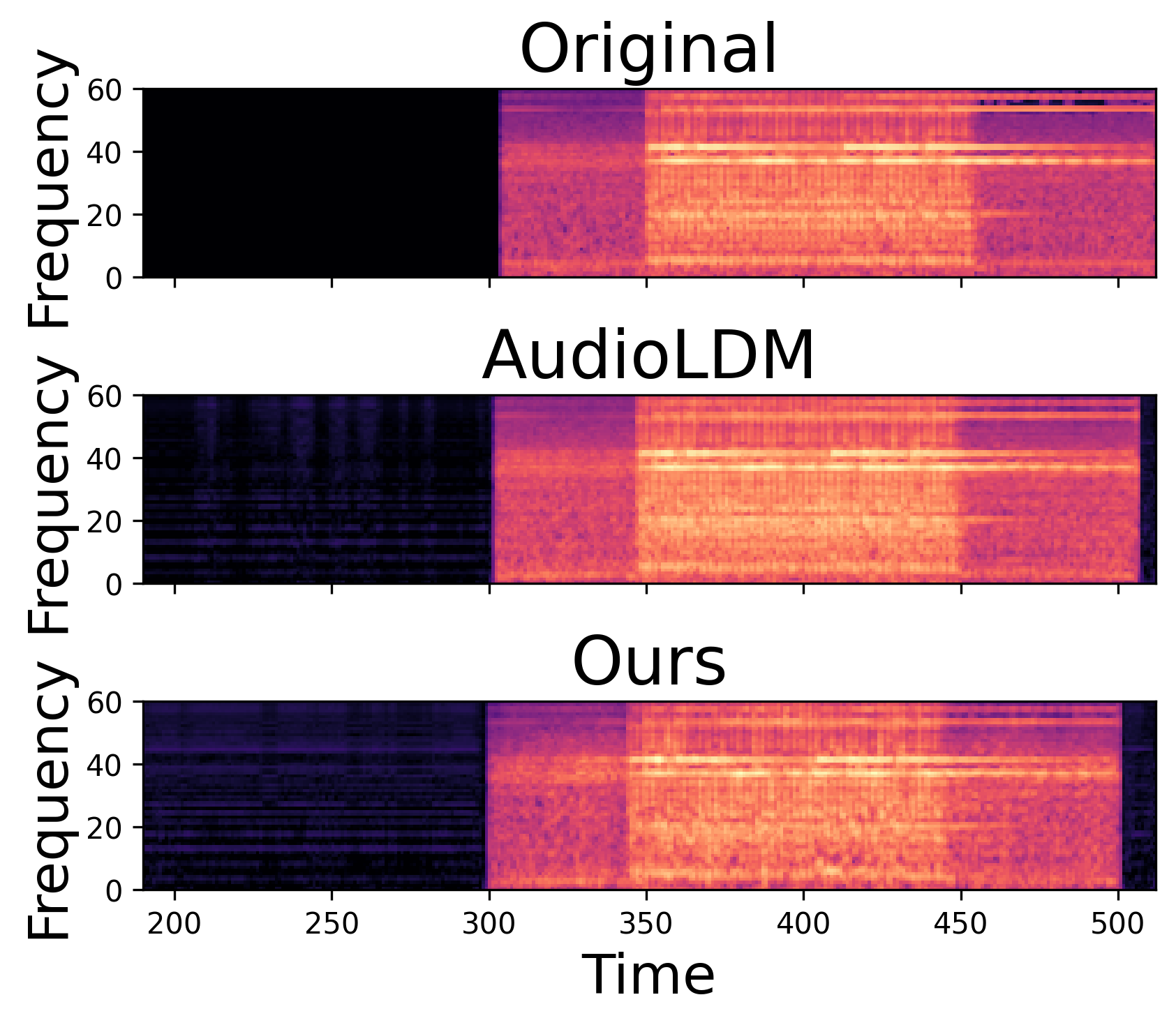} };
        \node [draw=none, fill=none, right of=ex1, xshift=4cm] (ex2)  { \includegraphics[width=0.25\textwidth]{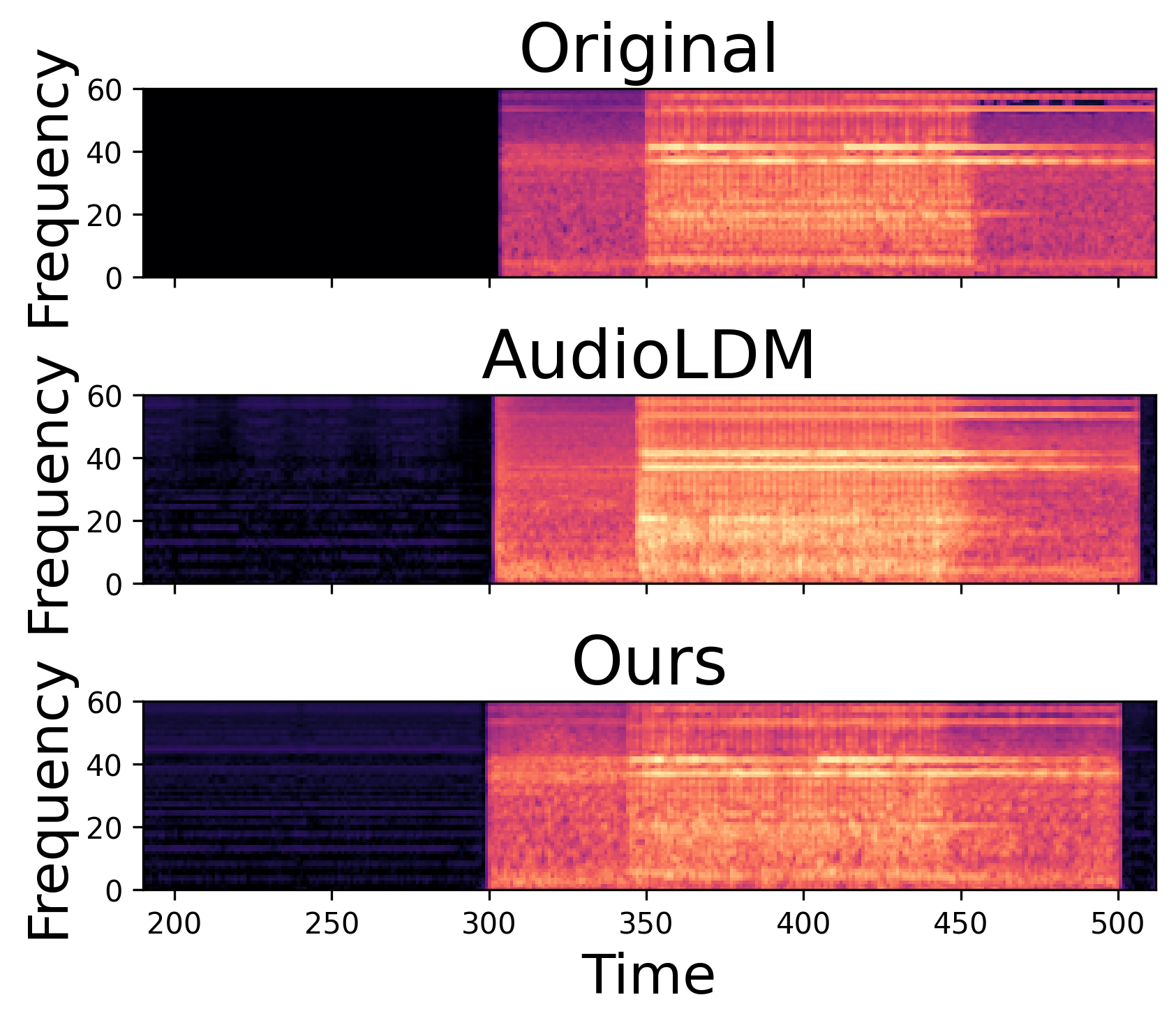} };
        \node [draw=none, fill=none, right of=ex2, xshift=4cm] (ex3)  { \includegraphics[width=0.25\textwidth]{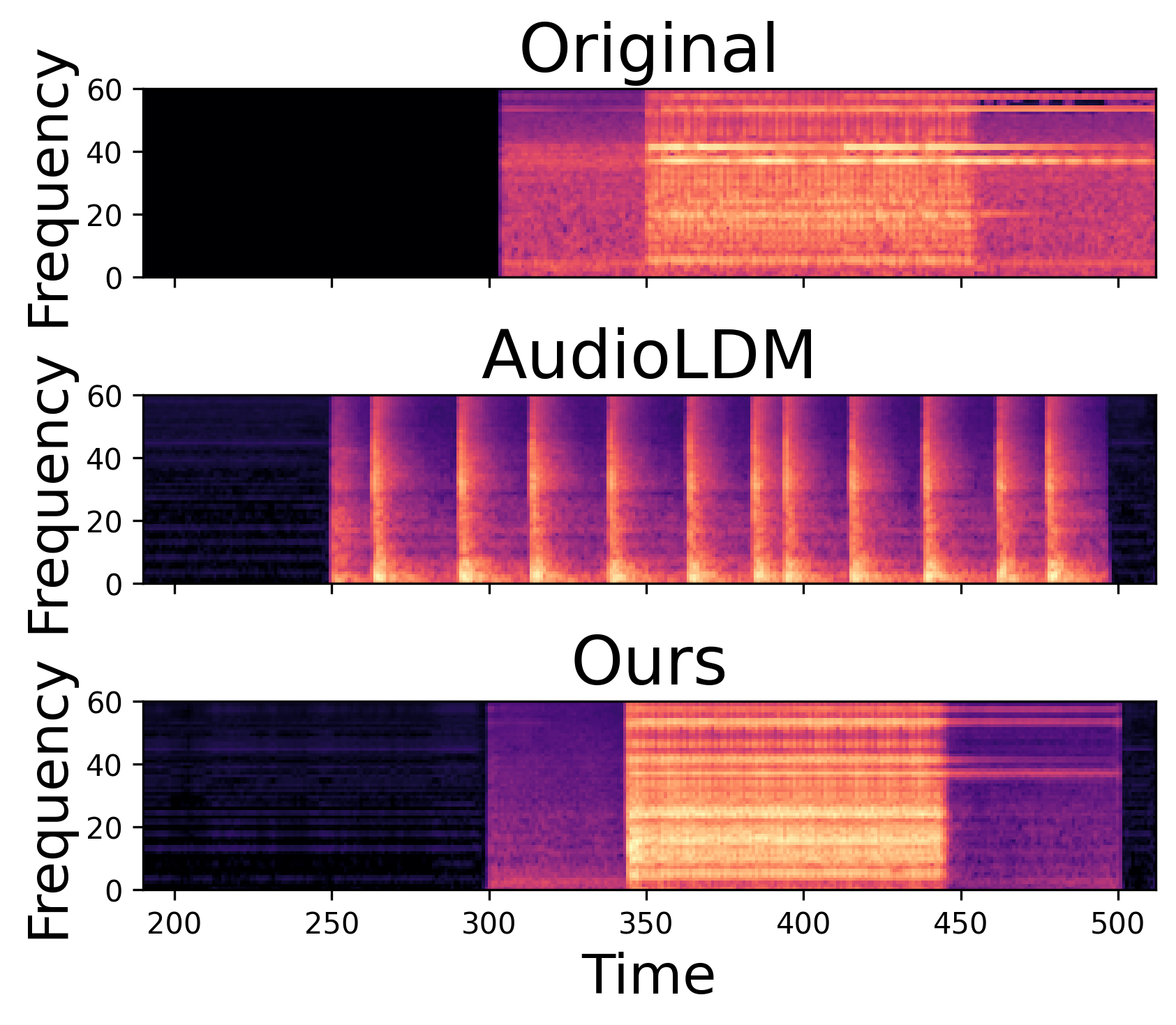} };
        \node [draw=none, fill=none, right of=ex3, xshift=4cm] (ex4)  { \includegraphics[width=0.25\textwidth]{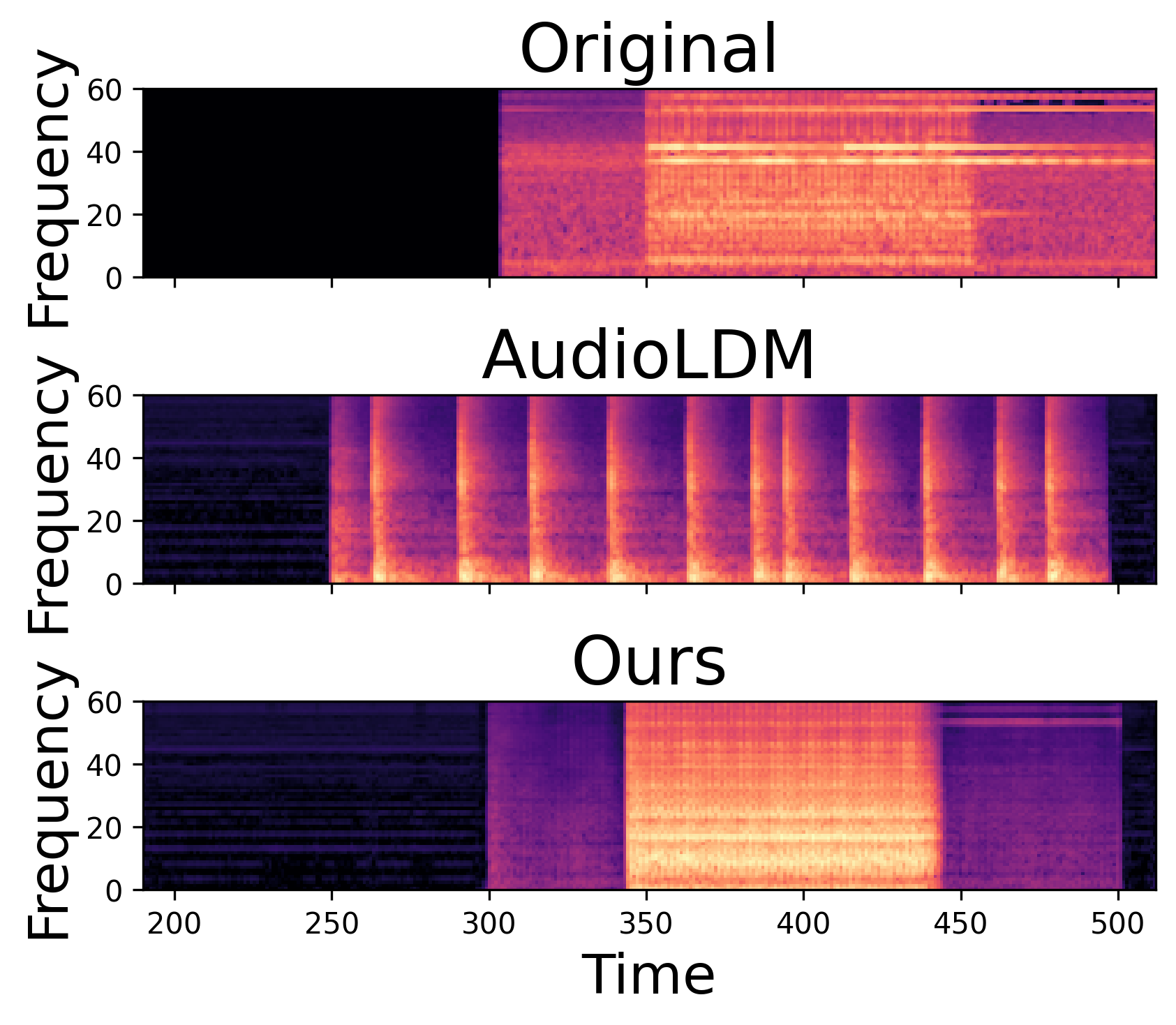} };
        \node [draw=none, fill=none, right of=ex4, xshift=4cm] (ex5)  { \includegraphics[width=0.25\textwidth]{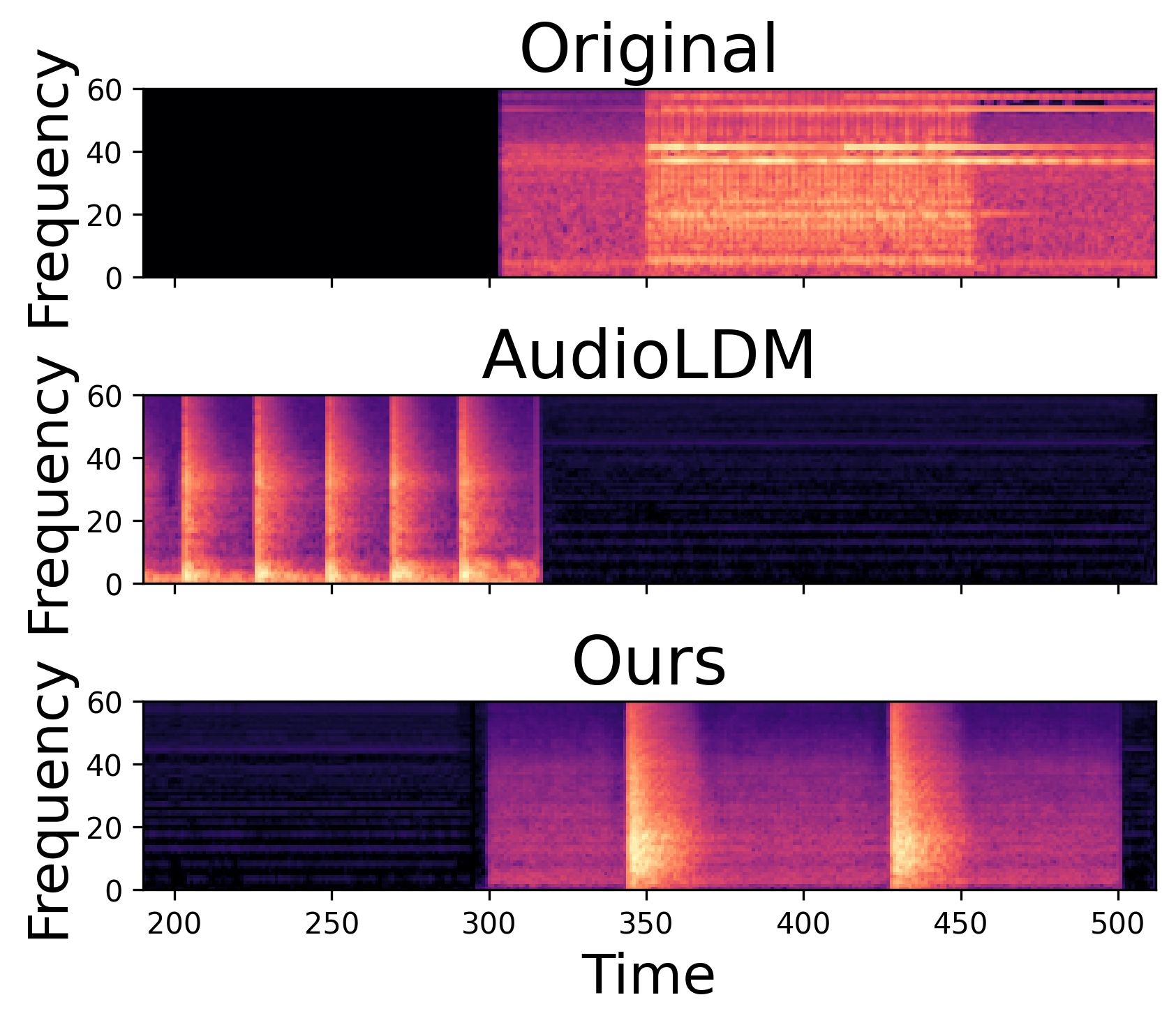} };

        \node [draw=none, fill=none, above of=ex1, yshift=1cm] (start)  { };
        \node [draw=none, fill=none, above of=ex4, xshift=4.5cm, yshift=1cm] (end)  { };
        \draw[->, thick, scale=0.7] (start.north west) -- (end.north east) node[midway, above] {Edit strength ($\eta$ for our pipeline, transfer strength for AudioLDM)};
        
    \end{tikzpicture}
    }
    \vspace{-0.3cm}
    \caption{Example showcasing a style transfer experiment. The input audio that corresponds to `phone ringing' is on the top row, and the text describing the target edit is `Sound of knocking on the door'. Our edited audio (bottom row) maintains the same onset pattern as the original sample better than AudioLDM (middle row). From left to right, we increased the edit strength to showcase qualitatively the patterns highlighted in Fig. ~\ref{fig:results}.}
    \vspace{-0.3cm}
    \label{fig:eta}
\end{figure*}
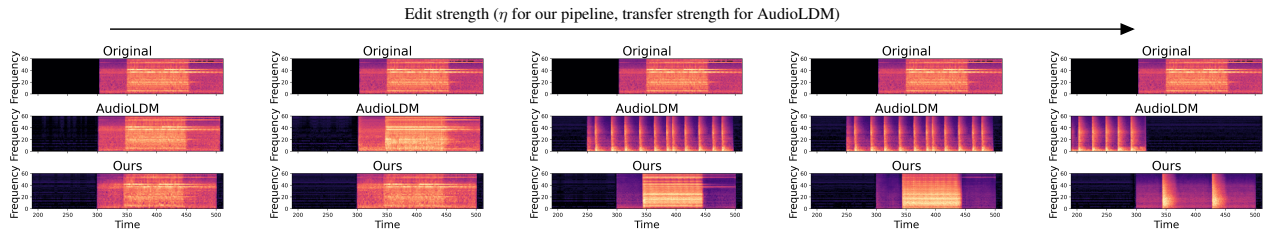

\section{Experimental setup}
\label{sec:setup}
To validate our audio editing methodology, we benchmark it on three edit types: addition, style transfer, and inpainting.
Since the proposed method is a non-rigid text-prompted editing pipeline, we want to emphasize that it supports free-form text descriptions of the target edits. However, we have limited ourselves to these three classes of edits to ascertain that the model can stay faithful to the input recordings. We also want to emphasize that these are the most accessible edit types to {be able to} evaluate and compare with other baselines, namely AUDIT, SDEdit, and AudioLDM. As expected from the explanation in Sec.~\ref{sec:methods}, the effect of the hyperparameters is similar among the three edit types. In particular, increasing the timesteps for the Latent Diffusion Model used in the editing pipeline reduces the fidelity to the input but improves the alignment with the target text. A similar trend is observed by increasing the $\eta$ values, as showcased in Fig. \ref{fig:eta}.

\subsection{Edit types}
In this subsection, we present the three edit types we experimented with in this paper. \\
\noindent \textbf{Addition.} This edit consists of adding a sound corresponding to the target text to the input sound while keeping fidelity to the latter.
In our approach, after performing the embedding optimization and diffusion fine-tuning, we can control the strength of the added signal with both the $\eta$ value and the number of diffusion timesteps, thus controlling their direct relevance to the target text. In particular, an increased number of timesteps gives more relevance to the target audio, i.e. the edit, but loses fidelity to the input. A similar trend is observed by increasing $\eta$ values. For this edit, similarly to Imagic \cite{kawar2023imagic}, the edits are described by giving the entire context. For example, to add gunshots to an audio sample containing sirens, the text prompt should look similar to `\textit{Sound of sirens wailing with gunshots in the background.}'\\
\textbf{Style transfer.} 
Style transfer is a general term for a type of edit where the style of the input audio is altered. For instance, a dog barking sound could be turned into a series of gunshots. An important thing to consider is that, when the end goal is `editing', the model needs to be able to keep a certain level of faithfulness to the input audio. 
More specifically, the output audio should have the same content (e.g. onsets, temporal occurrence) as the original audio while modifying the frequency content of the acoustic events. Fig.~\ref{fig:eta} compares a style transfer edit performed with our method and AudioLDM. Similarly to what we need to do for addition, the text should describe the event, and the editing pipeline will generate an aligned edit, depending on the interpolation parameters chosen. For example, to edit dog barking into the style of gunshots, the prompt for the edit should be similar to for instance `\textit{Sounds of gunshots.}' \\
\textbf{Inpainting.} We considered inpainting an edit, as it can be considered an addition with temporal occurrence taken into account. We want to fill a blank portion in the input audio during this edit. As in the previous cases, the edit added to the original audio should align with the original audio and the text prompt. In this case, the edit should describe what should be added in the empty portion of the input audio. For example, to complete the audio of a dog barking with dog barks, the edit prompt should be `\textit{Sound of a dog barking.}' This is the only edit type that does not require computing fidelity, as the blank portion of the audio does not have a ground truth and thus neither a target audio to be aligned to. To be precise, we only use the generated portion of the spectrogram that corresponds to the missing portion of the input audio, i.e. we do not replace the portions of the data that correspond to the observed input data. This ensures maximal fidelity with the generated edit. 

\subsection{Quantitative metrics}
\label{sec:metrics}
Similarly to \cite{kawar2023imagic}, which proposes an editing pipeline for images, we consider an edit good when it can balance the trade-off between preserving the audio content and being aligned with the target text. Quantitatively, we can use two metrics to measure these qualities, namely, the CLAP \cite{clap} score for audio and text. To compute the Text CLAP score, we project the edited audio and the text describing the edit into the shared multimodal latent space of CLAP, resulting in $\mathbf{\widehat h}_a$ and $\mathbf{\widehat h}_t$. Then, we compute the cosine similarity between the two embeddings:
\begin{equation}
    \text{Text CLAP}=\frac{\mathbf{\widehat h}_a^T\mathbf{h}_t}{||\mathbf{\widehat h}_a||\cdot||\mathbf{h}_t||}
\end{equation}
Similarly, we define the Audio CLAP score as the cosine similarity between $\mathbf{\widehat h}_a$ and the projection of the original audio in the shared multimodal latent space $\mathbf{h}_a$:
\begin{equation}
    \text{Audio CLAP}=\frac{\mathbf{\widehat h}_a^T\mathbf{h}_a}{||\mathbf{\widehat h}_a||\cdot||\mathbf{h}_a||}
\end{equation}
Higher CLAP scores correlate to better alignment with the text edit and higher fidelity to the input audio. To optimize the tradeoff between fidelity to the input and to the edit type, we propose to select the $\eta$ value that maximises the sum of CLAP scores, also reported in Table \ref{tab:results}. This allows selecting the interpolation value that yields the most faithful edit with respect to both the input audio and input text. 

\subsection{Text-to-audio model and optimization parameters}

\begin{figure*}[t]
    \centering
    \includegraphics[width=0.33\textwidth, trim=0.2cm 0cm 0.5cm 0cm,clip]{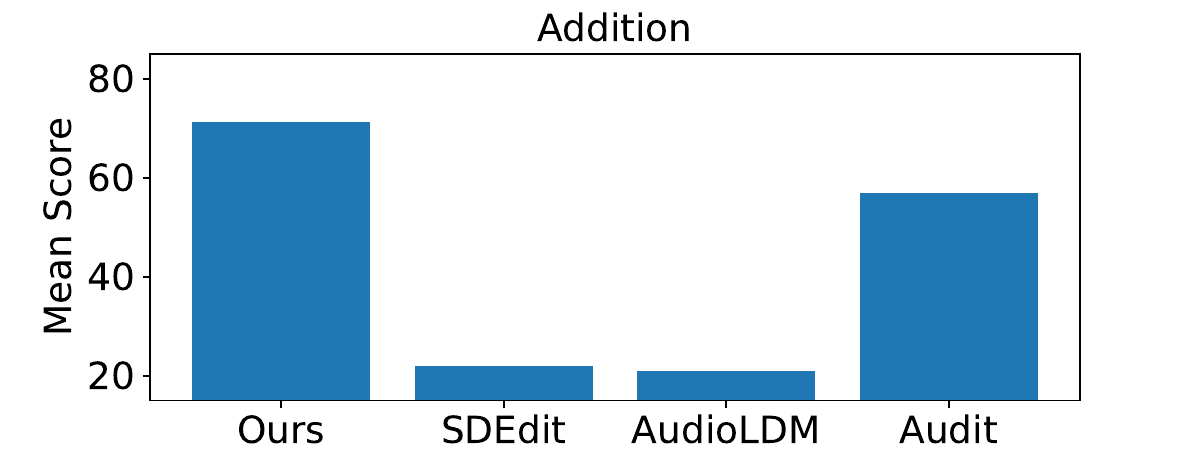}
    \includegraphics[width=0.33\textwidth,trim=.2cm 0cm 0.5cm 0.0cm,clip ]{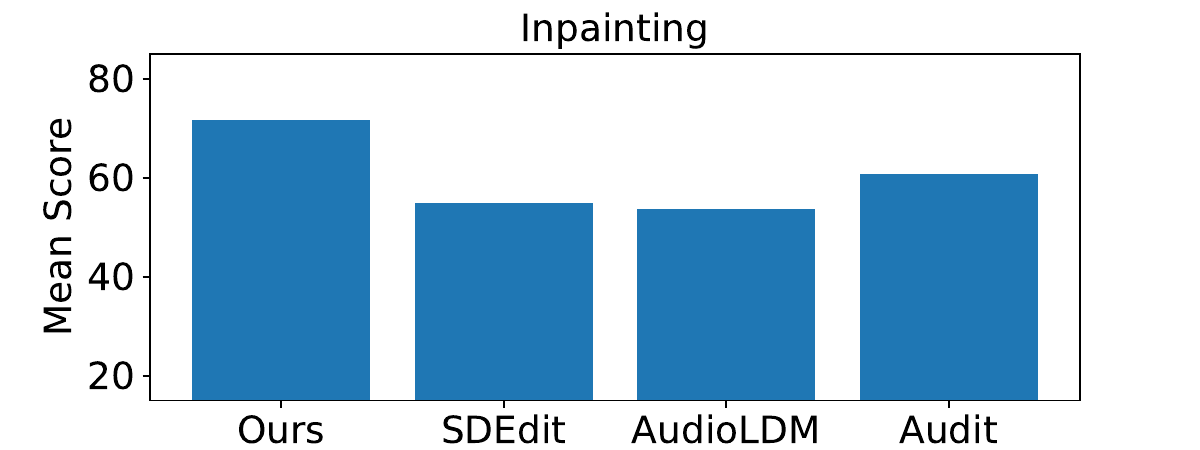}
    \includegraphics[width=0.33\textwidth, trim=.2cm 0cm 0.5cm 0.0cm,clip]{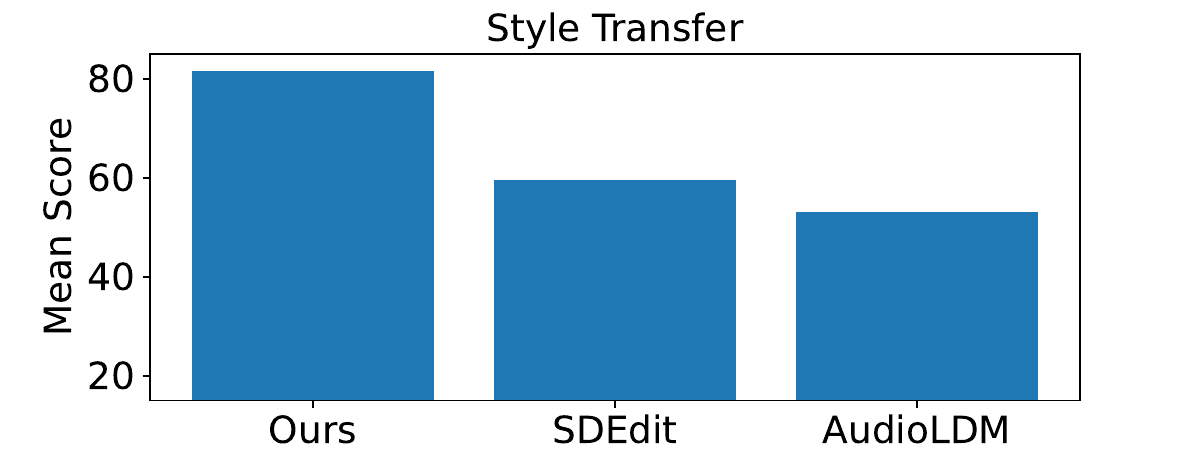}
    \vspace{-0.4cm}
    \caption{Comparisons of user preferences on \textbf{Addition}, \textbf{Inpainting} and \textbf{Style Transfer} tasks}
    \label{fig:userstudy}
    \vspace{-0.4cm}
\end{figure*}

\textbf{Generative model.} As the base generative model for the editing pipeline, we used the state-of-the-art TANGO model \cite{ghosal2023tango}, which leverages FLAN-T5~\cite{pmlr-v202-longpre23a} as a text encoder. However, note that the proposed approach can work on all the latent-diffusion-based generative models, such as AudioLDM \cite{liu2023audioldm} and AudioLDM 2 \cite{Liu2023AudioLDM2L}.\\
\textbf{Optimization process.} To perform Step 1 of the editing pipeline, described in Sec.~\ref{sec:methods}, we used the Adam~\cite{kingma2015adam} optimizer with a learning rate of \num{2e-3}, and performed \num{500} embedding optimization steps. Increasing the learning rate or the number of steps improves the overall alignment of the generated sample to the original one. However, this can compromise the interpolation, as $\mathbf e_\text{opt}$ and $\mathbf e_\text{tgt}$ could be too far from each other to work under the approximation of linear interpolation. For Step 2, we again used the Adam optimizer with a learning rate of \num{1e-6} and \num{1500} steps. In this case, increasing the learning rate or the number of optimization steps improves the alignment with the original sample but compromises the ability of the diffusion model to generate other sounds when conditioned with the $\mathbf e_\text{text}$ embedding. In fact, the model overfits the original sample and ignores text prompts. 
\\ \textbf{Editing speedup.} While running the embedding optimization step is relatively fast (0.42 s/step), the optimization fine-tuning process can take several minutes to complete, depending on the audio length. For example, it takes around 17 minutes on an NVIDIA RTX3090 GPU for a 10-second long audio. To reduce the time needed for the editing pipeline, we propose to use some techniques for Parameter-Efficent Fine-Tuning (PEFT)~\cite{ding2023parameter}, namely LoRA~\cite{Hu2021LoRALA}, to make Step 2 of the pipeline faster.

        

\section{Results}
To validate the quality of the proposed editing pipeline, we benchmarked it against AudioLDM~\cite{liu2023audioldm}, AUDIT~\cite{wang2023audit}, and SDEdit~\cite{Meng2021SDEditGI} both quantitatively and qualitatively. The code to reproduce the results is available on our companion website\footnote{\url{https://fpaissan.github.io/editingweb/}}.


\subsection{Qualitative comparison}

We presented 17 users (who are machine learning researchers) with samples from the AUDIT companion website, as the authors of AUDIT did not provide an implementation\footnote{\url{https://audit-demo.github.io/}}. The user study results are reported in Fig.~\ref{fig:userstudy}. Note that we used three samples for addition and inpainting, and two for style transfer. We asked the users to provide an opinion between 0-100 using the webMushra interface\footnote{\url{https://github.com/audiolabs/webMUSHRA}}. Our proposed method is preferred for the addition tasks over AudioLDM and SDEdit with a huge margin, while AUDIT is more comparable. Note that even though AUDIT works only with pre-specified edits, we improve on its user preference while maintaining flexibility on the edit instructions, that is our method support non-rigid text prompts while AUDIT can not. We observe that our method outperforms the other methods for the inpainting task. Finally, we observe that we obtained similar user performance on style transfer compared to SDEdit and AudioLDM, while we could not compare with AUDIT since it cannot perform this edit type.

\begin{table}[t]
\centering
\caption{Comparison of CLAP scores between our editing pipeline and the baselines. For all scores, higher is better. The results for our approach are generated with 200 diffusion timesteps, aligned with the suggested range from TANGO~\cite{ghosal2023tango}. In the table, the sum of CLAP scores is reported to showcase the tradeoff between fidelity to the input and fidelity to the edit.}
\begin{tabular}{c | l | l c c c }
    \toprule
    \textbf{Edit type} & \textbf{Edit pipeline} & \textbf{Sum} & \textbf{Text} & \textbf{Audio} \\ 
    \midrule
    \multirow{5}{*}{\rotatebox[origin=c]{90}{Addition}} 
     & AudioLDM & 1.229 & 0.524 & 0.705 \\
     & AUDIT & 0.985 & 0.208 & 0.777 \\
     & SDEdit & 0.925 & 0.237 & 0.688 \\
     & Ours & 1.366 & 0.544 & 0.822 \\
     & \textbf{Ours (LORA)} & \textbf{1.407} & 0.524 & 0.883 \\ 
    \midrule
    \multirow{5}{*}{\rotatebox[origin=c]{90}{Inpainting}} 
     & AudioLDM & 1.465 & 0.544 & 0.921 \\
     & AUDIT & 0.912 & 0.223 & 0.689 \\
     & SDEdit & 0.979 & 0.280 & 0.699 \\
     & Ours & 1.472 & 0.633 & 0.839 \\
     & \textbf{Ours (LORA)} & \textbf{1.499} & 0.645 & 0.854 \\ 
    \midrule
    \multirow{4}{*}{\rotatebox[origin=c]{90}{\parbox{1cm}{\centering Style \\ Transfer}}}
     & AudioLDM & 0.970 & 0.260 & 0.710 \\
     & SDEdit & 0.651 & 0.156 & 0.495 \\
     & Ours & 1.141 & 0.460 & 0.681 \\
     & \textbf{Ours (LORA)} & \textbf{1.161} & 0.465 & 0.696 \\ \bottomrule
\end{tabular}
\label{tab:results}
\end{table}

\subsection{Quantitative evaluation}

\begin{figure}[t]
    \centering
    \begin{subfigure}{.23\textwidth}
      \includegraphics[width=1.0\linewidth, trim=0cm 3.8cm 0cm 2cm]{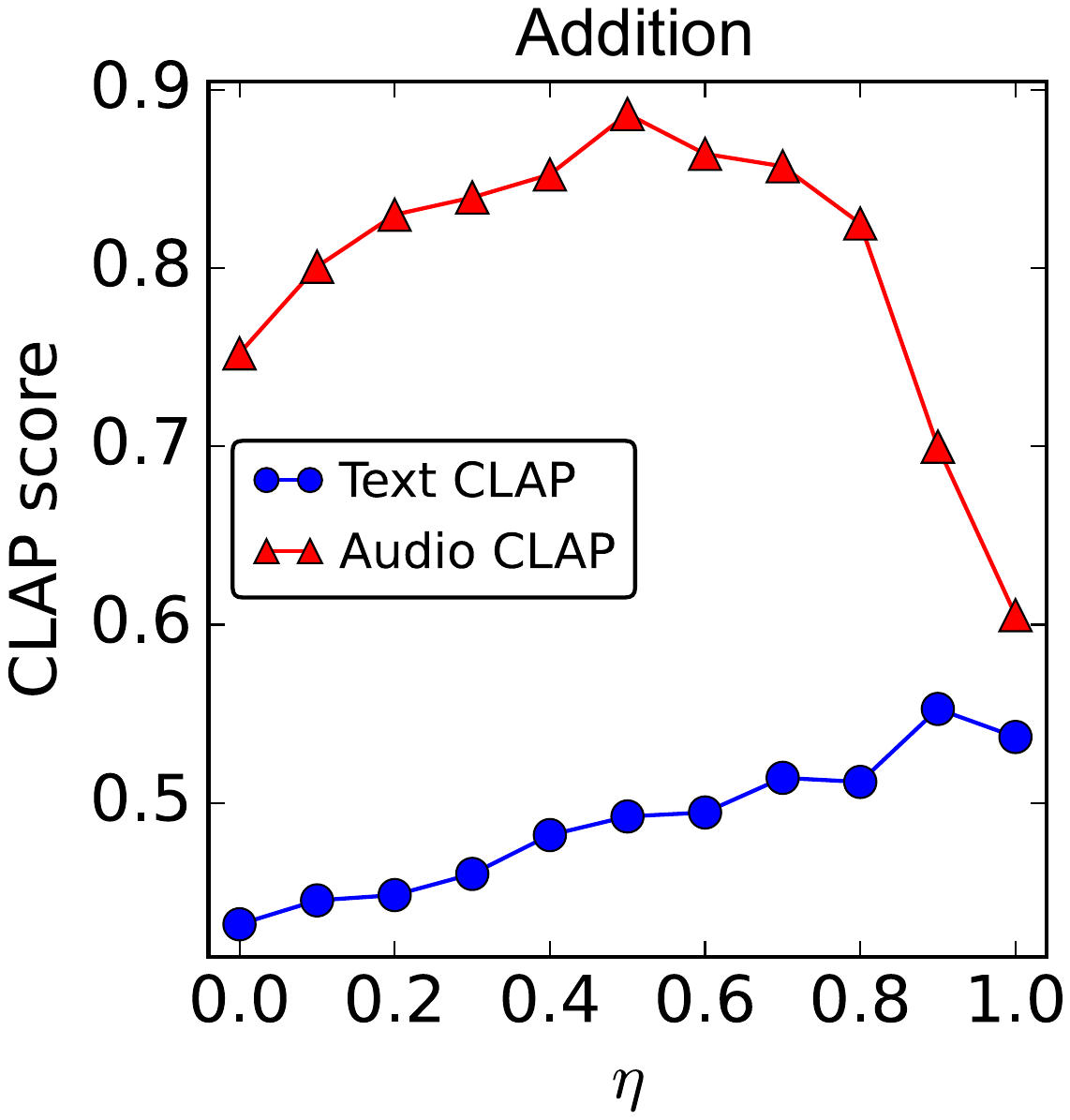}
    \end{subfigure}
    \begin{subfigure}{.23\textwidth}
      \includegraphics[width=1.0\linewidth, trim=0cm 3.8cm 0cm 2cm]{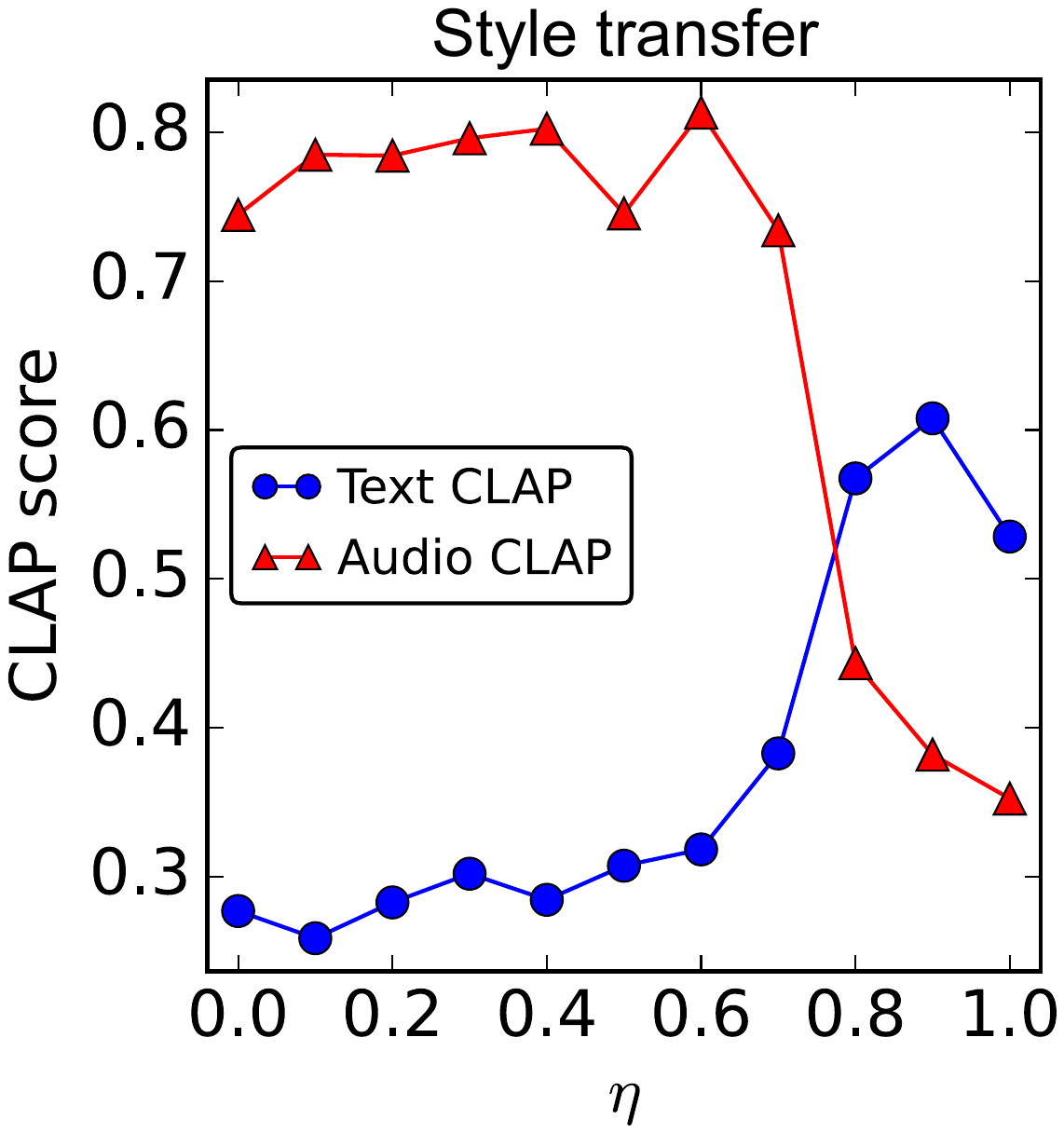}
    \end{subfigure}
    \vspace{-.3cm}
\caption{
Quantitative evaluation of the addition and style transfer edits over {9} samples for each type. The trends are in line with the intuition of $\eta$ as a parameter to control the edit strength.
}
\vspace{-0.65cm}
\label{fig:results}
\end{figure}

\vspace{-0.2cm}
While comparing different editing pipelines, we considered two aspects: how well the edits aligned with the target text and how faithful they were to the input audio. In Table \ref{tab:results}, we also report the sum of the two CLAP scores to showcase the tradeoff between these two aspects of audio edits. Reported results are obtained on twenty-seven \SI{10}{\s} long audio samples of acoustic scenes, each with an associated edit prompt. Some edits are showcased and can be listened to on the companion website\footnote{\url{https://fpaissan.github.io/editingweb/}}
\textbf{Fidelity to the input.}.
Audio CLAP scores compute the similarity of the edits to the input audio. As expected, increasing the edit strength decreases the similarity to the input audio, as can be observed in Fig.~\ref{fig:results} and Table \ref{tab:results}, and in line with the similar approach, Imagic \cite{kawar2023imagic}. Nonetheless, we observed that Audio CLAP is not well suited for comparing the onsets and offsets of the events. An example is shown in Fig.~\ref{fig:eta}, where we compare AudioLDM and our editing pipeline on the task of style transfer for different editing strengths. Our approach remains more aligned (faithful) to the onsets and offsets in the original audio compared to AudioLDM. Even for significantly large values of edit strength and, even though we get marginally lower scores of Audio CLAP scores, the edits generated by our approach are temporally located in the same region in which the original event happens, and present a good alignment with the edit prompt. \\
\textbf{Semantic alignment with target text.}
The other important quality of good edits is the alignment with the target text. As expected from Eq.~ \ref{eq:interpolate}, higher $\eta$ values translate to higher alignment, as proven is Fig.~ \ref{fig:results}. 
We should emphasize, however, that the CLAP score metric is slightly biased towards CLAP-based approaches such as AudioLDM and AUDIT as they are explicitly trained to optimize the metric.\\
\textbf{Impact of LoRA.} As expected from recent literature~\cite{fu2023effectiveness, dettmers2024qlora, li2024loftq}, also in this case, LoRA does not harm performance on the generated edits. In fact, using LoRA, we achieve the best tradeoff between edit and audio fidelity with almost half the time of the original editing pipeline (i.e. from 17 to 9 mins for a 10-second long audio on NVIDIA RTX3090 GPU).

\section{Conclusions}
In this paper, we have proposed to adapt the editing method presented in \cite{kawar2023imagic} for audio latent diffusion models. The experimental results indicate that the edits produced by our method outperform current state-of-the-art approaches for audio editing in terms of both quantitative and qualitative assessments. One limitation of the presented approach is that it requires optimization for each sample. To address this issue, we proposed to use LoRA in the pipeline to significantly speed up the editing speed without affecting the overall edit quality.


\bibliographystyle{IEEEtran}
\bibliography{mybib}

\begin{thebibliography}{10}
\providecommand{\url}[1]{#1}
\csname url@samestyle\endcsname
\providecommand{\newblock}{\relax}
\providecommand{\bibinfo}[2]{#2}
\providecommand{\BIBentrySTDinterwordspacing}{\spaceskip=0pt\relax}
\providecommand{\BIBentryALTinterwordstretchfactor}{4}
\providecommand{\BIBentryALTinterwordspacing}{\spaceskip=\fontdimen2\font plus
\BIBentryALTinterwordstretchfactor\fontdimen3\font minus \fontdimen4\font\relax}
\providecommand{\BIBforeignlanguage}[2]{{%
\expandafter\ifx\csname l@#1\endcsname\relax
\typeout{** WARNING: IEEEtran.bst: No hyphenation pattern has been}%
\typeout{** loaded for the language `#1'. Using the pattern for}%
\typeout{** the default language instead.}%
\else
\language=\csname l@#1\endcsname
\fi
#2}}
\providecommand{\BIBdecl}{\relax}
\BIBdecl

\bibitem{yang2023diffusion}
L.~Yang, Z.~Zhang, Y.~Song, S.~Hong, R.~Xu, Y.~Zhao, W.~Zhang, B.~Cui, and M.-H. Yang, ``Diffusion models: A comprehensive survey of methods and applications,'' \emph{ACM Computing Surveys}, vol.~56, no.~4, pp. 1--39, 2023.

\bibitem{liu2024visual}
H.~Liu, C.~Li, Q.~Wu, and Y.~J. Lee, ``Visual instruction tuning,'' in \emph{International Conference on Neural Information Processing Systems (NeurIPS)}, vol.~36, 2024.

\bibitem{li2024snapfusion}
Y.~Li, H.~Wang, Q.~Jin, J.~Hu, P.~Chemerys, Y.~Fu, Y.~Wang, S.~Tulyakov, and J.~Ren, ``{SnapFusion}: Text-to-image diffusion model on mobile devices within two seconds,'' in \emph{International Conference on Neural Information Processing Systems (NeurIPS)}, vol.~36, 2024.

\bibitem{Karras2017ProgressiveGO}
T.~Karras, T.~Aila, S.~Laine, and J.~Lehtinen, ``Progressive growing of {GAN}s for improved quality, stability, and variation,'' in \emph{International Conference on Learning Representations (ICLR)}, 2018.

\bibitem{Ramesh2021ZeroShotTG}
A.~Ramesh, M.~Pavlov, G.~Goh, S.~Gray, C.~Voss, A.~Radford, M.~Chen, and I.~Sutskever, ``Zero-shot text-to-image generation,'' in \emph{International Conference on Machine Learning (ICML)}, vol. 139, 2021, pp. 8821--8831.

\bibitem{Rombach2021HighResolutionIS}
R.~Rombach, A.~Blattmann, D.~Lorenz, P.~Esser, and B.~Ommer, ``High-resolution image synthesis with latent diffusion models,'' in \emph{IEEE/CVF Conference on Computer Vision and Pattern Recognition (CVPR)}, 2021, pp. 10\,674--10\,685.

\bibitem{Brooks2022InstructPix2PixLT}
T.~Brooks, A.~Holynski, and A.~A. Efros, ``{InstructPix2Pix}: Learning to follow image editing instructions,'' in \emph{IEEE/CVF Conference on Computer Vision and Pattern Recognition (CVPR)}, 2022, pp. 18\,392--18\,402.

\bibitem{Couairon2022DiffEditDS}
G.~Couairon, J.~Verbeek, H.~Schwenk, and M.~Cord, ``{DiffEdit}: Diffusion-based semantic image editing with mask guidance,'' in \emph{International Conference on Learning Representations (ICLR)}, 2023.

\bibitem{huang2023make}
R.~Huang, J.~Huang, D.~Yang, Y.~Ren, L.~Liu, M.~Li, Z.~Ye, J.~Liu, X.~Yin, and Z.~Zhao, ``{Make-An-Audio}: Text-to-audio generation with prompt-enhanced diffusion models,'' in \emph{International Conference on Machine Learning (ICML)}, 2023, pp. 13\,916--13\,932.

\bibitem{Liu2023AudioLDM2L}
H.~Liu, Q.~Tian, Y.~Yuan, X.~Liu, X.~Mei, Q.~Kong, Y.~Wang, W.~Wang, Y.~Wang, and M.~D. Plumbley, ``{AudioLDM 2}: Learning holistic audio generation with self-supervised pretraining,'' \emph{arXiv preprint arXiv:2308.05734}, 2023.

\bibitem{liu2023audioldm}
H.~Liu, Z.~Chen, Y.~Yuan, X.~Mei, X.~Liu, D.~Mandic, W.~Wang, and M.~D. Plumbley, ``{AudioLDM}: Text-to-audio generation with latent diffusion models,'' \emph{International Conference on Machine Learning (ICML)}, 2023.

\bibitem{clap}
B.~Elizalde, S.~Deshmukh, M.~Al~Ismail, and H.~Wang, ``{CLAP}: learning audio concepts from natural language supervision,'' in \emph{IEEE International Conference on Acoustics, Speech and Signal Processing (ICASSP)}, 2023, pp. 1--5.

\bibitem{kreuk2022audiogen}
F.~Kreuk, G.~Synnaeve, A.~Polyak, U.~Singer, A.~D{\'e}fossez, J.~Copet, D.~Parikh, Y.~Taigman, and Y.~Adi, ``{AudioGen}: Textually guided audio generation,'' in \emph{International Conference on Learning Representations (ICLR)}, 2023.

\bibitem{copet2024simple}
J.~Copet, F.~Kreuk, I.~Gat, T.~Remez, D.~Kant, G.~Synnaeve, Y.~Adi, and A.~D{\'e}fossez, ``Simple and controllable music generation,'' in \emph{International Conference on Neural Information Processing Systems (NeurIPS)}, 2023.

\bibitem{ghosal2023tango}
D.~Ghosal, N.~Majumder, A.~Mehrish, and S.~Poria, ``Text-to-audio generation using instruction guided latent diffusion model,'' in \emph{International Conference on Multimedia}, 2023, pp. 3590–--3598.

\bibitem{wang2023audit}
Y.~Wang, Z.~Ju, X.~Tan, L.~He, Z.~Wu, J.~Bian, and sheng zhao, ``{AUDIT}: Audio editing by following instructions with latent diffusion models,'' in \emph{International Conference on Neural Information Processing Systems (NeurIPS)}, 2023.

\bibitem{Meng2021SDEditGI}
C.~Meng, Y.~He, Y.~Song, J.~Song, J.~Wu, J.-Y. Zhu, and S.~Ermon, ``{SDEdit}: Guided image synthesis and editing with stochastic differential equations,'' in \emph{International Conference on Learning Representations (ICLR)}, 2021.

\bibitem{kawar2023imagic}
B.~Kawar, S.~Zada, O.~Lang, O.~Tov, H.~Chang, T.~Dekel, I.~Mosseri, and M.~Irani, ``Imagic: Text-based real image editing with diffusion models,'' in \emph{IEEE/CVF Conference on Computer Vision and Pattern Recognition (CVPR)}, 2023, pp. 6007--6017.

\bibitem{Hu2021LoRALA}
E.~J. Hu, yelong shen, P.~Wallis, Z.~Allen-Zhu, Y.~Li, S.~Wang, L.~Wang, and W.~Chen, ``Lo{RA}: Low-rank adaptation of large language models,'' in \emph{International Conference on Learning Representations (ICLR)}, 2022.

\bibitem{Kingma2014}
D.~P. Kingma and M.~Welling, ``Auto-encoding variational {Bayes},'' in \emph{International Conference on Learning Representations (ICLR)}, 2014.

\bibitem{Kong2020HiFiGANGA}
J.~Kong, J.~Kim, and J.~Bae, ``{HiFi-GAN}: generative adversarial networks for efficient and high fidelity speech synthesis,'' in \emph{International Conference on Neural Information Processing Systems (NeurIPS)}, 2020.

\bibitem{ronneberger2015unet}
O.~Ronneberger, P.~Fischer, and T.~Brox, ``{U-Net}: Convolutional networks for biomedical image segmentation,'' in \emph{Medical Image Computing and Computer-Assisted Intervention (MICCAI)}, N.~Navab, J.~Hornegger, W.~M. Wells, and A.~F. Frangi, Eds., 2015, pp. 234--241.

\bibitem{pmlr-v202-longpre23a}
S.~Longpre, L.~Hou, T.~Vu, A.~Webson, H.~W. Chung, Y.~Tay, D.~Zhou, Q.~V. Le, B.~Zoph, J.~Wei, and A.~Roberts, ``{The Flan Collection}: Designing data and methods for effective instruction tuning,'' in \emph{International Conference on Machine Learning (ICML)}, vol. 202, 2023, pp. 22\,631--22\,648.

\bibitem{kingma2015adam}
D.~Kingma and J.~Ba, ``Adam: A method for stochastic optimization,'' in \emph{International Conference on Learning Representations (ICLR)}, 2015.

\bibitem{ding2023parameter}
N.~Ding, Y.~Qin, G.~Yang, F.~Wei, Z.~Yang, Y.~Su, S.~Hu, Y.~Chen, C.-M. Chan, W.~Chen \emph{et~al.}, ``Parameter-efficient fine-tuning of large-scale pre-trained language models,'' \emph{Nature Machine Intelligence}, vol.~5, no.~3, pp. 220--235, 2023.

\bibitem{fu2023effectiveness}
Z.~Fu, H.~Yang, A.~M.-C. So, W.~Lam, L.~Bing, and N.~Collier, ``On the effectiveness of parameter-efficient fine-tuning,'' in \emph{AAAI Conference on Artificial Intelligence}, vol.~37, no.~11, 2023, pp. 12\,799--12\,807.

\bibitem{dettmers2024qlora}
T.~Dettmers, A.~Pagnoni, A.~Holtzman, and L.~Zettlemoyer, ``{QL}o{RA}: Efficient finetuning of quantized {LLM}s,'' in \emph{International Conference on Neural Information Processing Systems (NeurIPS)}, 2023.

\bibitem{li2024loftq}
Y.~Li, Y.~Yu, C.~Liang, N.~Karampatziakis, P.~He, W.~Chen, and T.~Zhao, ``{LoftQ}: {L}o{RA}-fine-tuning-aware quantization for large language models,'' in \emph{International Conference on Learning Representations (ICLR)}, 2024.

\end{thebibliography}

\end{document}